\documentclass{elsart}

\usepackage{graphicx}

\usepackage{citesort,epsfig,enumerate,color}
\newcommand{\floor}[1]{\left\lfloor{#1}\right\rfloor}
\newcommand{\ceiling}[1]{\left\lceil{#1}\right\rceil}
\newcommand{\setof}[1]{\left\{{#1}\right\}}
\newcommand{\Xomit}[1]{}
\newtheorem{lemma}{Lemma}
\newtheorem{theorem}{Theorem}
\def\myendproof{{\ \vbox{\hrule\hbox{%
   \vrule height1.3ex\hskip0.8ex\vrule}\hrule }}\par}
\newenvironment{proof}{\noindent{\bf Proof. }}{\myendproof}

\begin{document}

\begin{frontmatter}



\title{Compact Floor-Planning via Orderly Spanning
Trees\thanksref{title}}

\thanks[title]{An early version of this work was presented at 9th
International Symposium on Graph Drawing, Vienna, Austria, September
2001.}

\author[address13]{Chien-Chih Liao}
\ead{henry@cobra.ee.ntu.edu.tw}
\author[address2]{Hsueh-I Lu\corauthref{cor1}}
\ead{hil@iis.sinica.edu.tw}
\ead[url]{http://www.iis.sinica.edu.tw/\~{ }hil/}
\corauth[cor1]{Coresponding author.}
\author[address13]{Hsu-Chun Yen\thanksref{author3}}
\thanks[author3]{Research of this author
was supported in part by NSC Grant 90-2213-E-002-100.}
\ead{yen@cc.ee.ntu.edu.tw}
\ead[url]{http://www.ee.ntu.edu.tw/\~{ }yen/}
\address[address13]{Department of Electrical Engineering, National
Taiwan University, Taipei 106, Taiwan, Republic of China.}
\address[address2]{Institute of Information Sicence, Academia Sinica,
Taipei 115, Taiwan, Republic of China.}


\begin{abstract}
{\em Floor-planning} is a fundamental step in VLSI chip design.  Based
upon the concept of {\em orderly spanning trees}, we present a simple
$O(n)$-time algorithm to construct a floor-plan for any $n$-node plane
triangulation. In comparison with previous floor-planning algorithms
in the literature, our solution is not only simpler in the algorithm
itself, but also produces floor-plans which require fewer module
types.  An equally important aspect of our new algorithm lies in its
ability to fit the floor-plan area in a rectangle of size $(n-1)\times
\floor{\frac{2n+1}{3}}$. Lower bounds on the worst-case area for
floor-planning any plane triangulation are also provided in the paper.
\end{abstract}


\end{frontmatter}

\section{Introduction}
\begin{figure}[t]
\centerline{\input{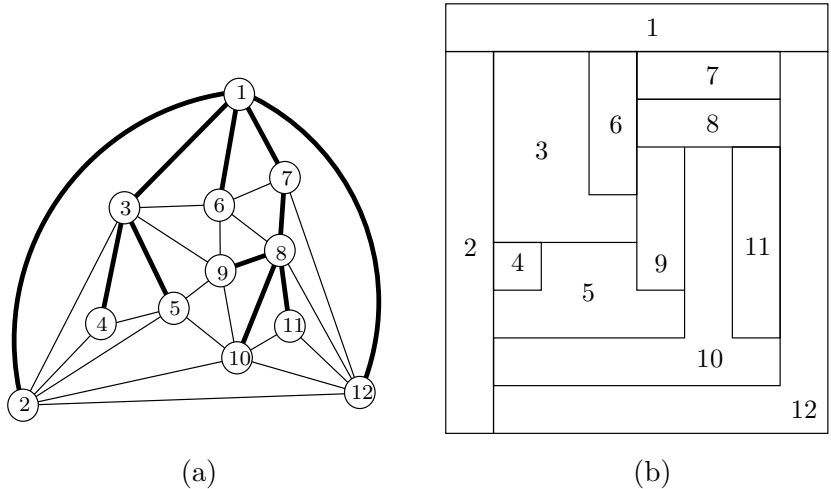}}
\caption{(a) A plane triangulation $G$, where an orderly spanning tree
$T$ of $G$ rooted at node 1 is drawn in dark. The node labels show the
counterclockwise preordering of the nodes in $T$. (b) A floor-plan of
$G$.}
\label{figure:example}
\end{figure}

In VLSI chip design, {\em
floor-planning}~\cite{TsukiyamaKS86,MailingMH82} refers to the process
of, given a graph whose nodes (respectively, edges) representing
functional entities (respectively, interconnections), partitioning a
rectangular chip area into a set of non-overlapping rectilinear
polygonal modules (each of which describes a functional entity) in
such a way that the modules of adjacent nodes share a common boundary.
For example, Figure~\ref{figure:example}(b) is a floor-plan of the
graph in Figure~\ref{figure:example}(a).

\begin{figure}[t]
\centerline{\input{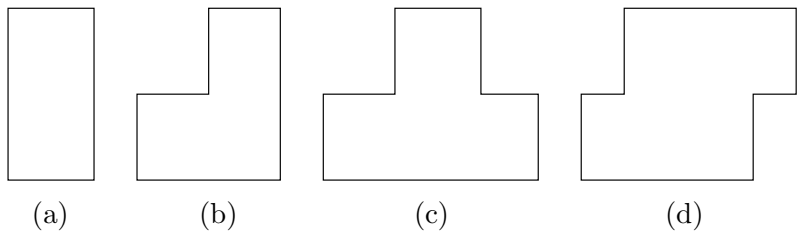}}
\caption{Four types of modules required by He's floor-planning
algorithm~\cite{He99}: (a) I-module, (b) L-module, (c) T-module, and
(d) Z-module. Our algorithm does not need Z-modules.}
\label{figure:module-type}
\end{figure}

Early stage of the {\em floor-planning} research focused on using {\em
rectangular modules} as the underlying building blocks.  A floor-plan
using only rectangles to represent nodes is called a {\em rectangular
dual}.  It was shown in~\cite{KozminskiK85,KozminskiK98,LaiL90} that a
plane triangulation $G$ admits a rectangular dual if and only if $G$
has four exterior nodes, and $G$ has no {\em separating triangles}.
(A separating triangle, which is also known as complex
triangle~\cite{YeapS93,TsukiyamaKS86}, is a cycle of three edges
enclosing some nodes in its interior.)  As for floor-planning general
plane graphs, Yeap and Sarrafzadeh~\cite{YeapS93} showed that
rectilinear modules with at most two concave corners are sufficient
and necessary.

In a subsequent study of floor-planning, He~\cite{He99} measured the
complexity of a module in terms of the number of its constituent
rectangles, as opposed to the number of concave corners.  A module
that is a union of $k$ or fewer disjoint rectangles is called a {\em
$k$-rectangular module}.  Since any rectilinear module with at most
two concave corners can be constructed by three rectangular modules,
the result of Yeap and Sarrafzadeh~\cite{YeapS93} implies the
feasibility of floor-planning plane graphs using 3-rectangular
modules.  He~\cite{He99} presented a linear-time algorithm to
construct a floor-plan of a plane triangulation using only
2-rectangular modules.  He's floor-planning algorithms consists of
three phases: The first phase utilizes the {\em canonical
ordering}~\cite{DeFPP90,Kant96,KantH97} to assign nodes on separating
triangles.  The second phase involves the so-called {\em vertex
expansion} operation to break all separating triangles. The third
phase adapts rectangular-dual
algorithms~\cite{BhaskerS87,BhaskerS88,He93,KantH97} to finalize the
drawing of the floor-plan.  Figure~\ref{figure:module-type} depicts
the shapes of the 2-rectangular modules required by He's algorithm.
For convenience, these four shapes are referred to as {\em I-module},
{\em L-module}, {\em T-module}, and {\em Z-module} throughout the rest
of this paper.

In this paper, we provide a ``simpler'' linear-time algorithm that
computes ``compact'' floor-plans for plane triangulations. The
``compactness'' of the output floor-plans is an important advantage of
our algorithm. 
Although previous work~\cite{He99,YeapS93}
reveals no area information, one can verify that a floor-plan 
using only $O(1)$-rectangular modules may require area 
$\Omega(n)\times \Omega(n)$.
The output of our algorithm for an $n$-node plane
triangulation has area no more than $(n-1) \times
\floor{\frac{2n+1}{3}}$, which can be shown to be
almost worst-case optimal.  What ``simplicity''
means is two-fold:
\begin{itemize}
\item First, as opposed to the multiple-phase approach
of~\cite{He99,YeapS93}, our algorithm is based upon a recent
development of {\em orderly spanning trees}~\cite{ChiangLL01}, which
provides an extension of {\em canonical
ordering}~\cite{DeFPP90,Kant96,KantH97} to plane graphs not required
to be triconnected and an extension for {\em
realizer}~\cite{Schnyder90,Schnyder89} to plane graphs not required to
be triangulated.  Our approach bypasses the somewhat complicated
rectangular-dual phase.  Aside from the two applications of orderly
spanning trees reported in~\cite{ChiangLL01} (namely, succinct
encodings for planar graphs with efficient query
support~\cite{Jacobson89,MR97,ChuangGHKL98} and 2-visibility drawings
for planar graphs~\cite{Foessmeierkk97}), our investigation here finds
another interesting application of orderly spanning trees.
(A similar concept called {\em ordered stratification} and its application in constructing 2-visibility drawing were independently studied by Bonichon, Le~Sa\"{e}c, and Mosbah~\cite{Bonichon00}.)  

\item Second, the floor-plan design of our algorithm is ``simpler''
(in comparison with~\cite{He99}) in its own right, in the sense that
I-modules, L-modules, and T-modules suffice.  (Recall that Z-modules
are needed by He's algorithm~\cite{He99}.) Our result is worst-case
optimal, since there is a plane triangulation that does not admit any
floor-plan consisting of only I-modules and
L-modules~\cite[Figure~4]{SunS93}.
\end{itemize}

The remainder of this paper is organized as follows.
Section~\ref{section:2} reviews the definition and property of orderly
spanning tree for plane graph.  Section~\ref{section:3} presents our
linear-time floor-planning algorithm as well as its correctness proof.
Section~\ref{section:4} provides a lower bound for the required area
for floor-planning plane triangulations.  Section~\ref{section:5}
concludes the paper.

\section{Orderly spanning tree}
\label{section:2}
A {\em plane graph} is a planar graph equipped with a fixed planar
embedding.  The embedding of a plane graph divides the plane into a
number of connected regions, each of which is called a {\em face}. The
unbounded face of $G$ is called the {\em exterior face}, whereas the
remaining faces are {\em interior faces}.  $G$ is a {\em plane
triangulation} if $G$ has at least three nodes and the boundary of
each face, including the exterior face, of $G$ is a triangle.  Let $T$
be a rooted spanning tree of a plane graph $G$. Two nodes are {\em
unrelated} in $T$ if they are distinct and neither of them is an
ancestor of the other in $T$. An edge of $G$ is {\em unrelated} with
respect to $T$ if its endpoints are unrelated in $T$. Let
$v_{1},v_{2},\ldots,v_{n}$ be the counterclockwise preordering of the
nodes in $T$. A node $v_{i}$ is {\em orderly} in $G$ with respect to
$T$ if the neighbors of $v_{i}$ in $G$ form the following four blocks
in counterclockwise order around $v_{i}$:
\begin{enumerate}[$B_1(v_{i})$:]
\item the parent of $v_{i}$,
\item the unrelated neighbors $v_j$ of $v_i$ with $j<i$,
\item the children of $v_{i}$, and
\item the unrelated neighbors $v_j$ of $v_i$ with $j>i$,
\end{enumerate}
where each block could be empty.  $T$ is an {\em orderly spanning
tree} of $G$ if $v_{1}$ is on the boundary of $G$'s exterior face, and
each $v_{i}, 1 \leq i \leq n$, is orderly in $G$ with respect to $T$.
It is not difficult to see that if $G$ is a plane triangulation, then
$B_2(v_i)$ (respectively, $B_4(v_i)$) is nonempty for each
$i=3,4,\ldots,n$ (respectively, $i=2,3,\ldots,n-1$). For each
$i=2,3,\ldots,n$, let $p(i)$ be the index of the parent of $v_i$ in
$T$.  Let $w(i)$ denote the number of leaves in the subtree of $T$
rooted at $v_i$.  Let $\ell(i)$ and $r(i)$ be the functions such that
$v_{\ell(i)}$ (respectively, $v_{r(i)}$) is the last (respectively,
first) neighbor of $v_i$ in $B_2(v_i)$ (respectively, $B_4(v_i)$) in
counterclockwise order around $v_i$. For example, in the example shown
in Figure~\ref{figure:example}(a), one can easily verify that node 3
is indeed orderly with respect to $T$, where $B_1(3)=\setof{1}$,
$B_2(3)=\setof{2}$, $B_3(3)=\setof{4,5}$, $B_4(3)=\setof{6,9}$,
$p(3)=1$, $w(3)=2$, $\ell(3)=2$, and $r(3)=9$. When $G$ is a plane
triangulation, it is known~\cite{ChiangLL01} that for each edge
$(v_i,v_j)$ of $G-T$ with $i<j$, at least one of $i=\ell(j)$ and
$j=r(i)$ holds. To be more specific, if $i=2$ and $j=n$, then both
$2=\ell(n)$ and $n=r(2)$ hold; otherwise, precisely one of $i=\ell(j)$
and $j=r(i)$ holds.

The concept of orderly spanning tree for plane
graphs~\cite{ChiangLL01} extends that of {\em canonical
ordering}~\cite{DeFPP90,Kant96,KantH97} for plane graphs not required
to be triconnected and that of {\em
realizer}~\cite{Schnyder90,Schnyder89,DeFraysseixOR94} for plane graphs not required
to be triangulated. 
Specifically, when $G$ is a plane triangulation,
(i) if $T$ is an orderly spanning tree of $G$, then the
counterclockwise preordering of the nodes of $T$ is always a canonical
ordering of $G$, and (ii) if $(T_1,T_2,T_n)$ is a realizer of $G$,
where $T_i$ is rooted at $v_i$ for each $i=1,2,n$, then each $T_i$
plus both external edges of $G$ incident to $v_i$ is an orderly
spanning tree of $G$.  Our floor-planning algorithm is based upon the
following lemma.
\begin{lemma}[see~\cite{ChiangLL01}]
\label{lemma:orderly}
Given an $n$-node plane triangulation $G$, an orderly spanning tree
$T$ of $G$ with at most $\floor{\frac{2n+1}{3}}$ leaves is obtainable
in $O(n)$ time.
\end{lemma}

\section{Our floor-planning algorithm}
\label{section:3}
A {\em floor-plan} $F$ of $G$ is a partition of a rectangle into
$n$ non-overlapping rectangular modules $r_1,r_2,\ldots,r_n$ such
that $v_i$ and $v_j$ are adjacent in $G$ if and only if the
boundaries of $r_i$ and $r_j$ share at least one non-degenerated
line segment.  The {\em size} of $F$ is the area of the rectangle
being partitioned by $F$ with the convention that the corners of
all modules are placed on integral grid points. For example, the
size of the floor-plan shown in Figure~\ref{figure:example}(b) is
$9\times 8$.
This section proves the following main theorem of the paper.
\begin{theorem}
\label{theorem:main}
Given an $n$-node plane triangulation $G$ with $n\geq 3$,
a floor-plan $F$ of $G$ can be constructed in $O(n)$ time such that
\begin{enumerate}
\item $F$ consists of I-modules, L-modules, and T-modules only, and
\item the size of $F$ is bounded by $(n-1)\times \floor{\frac{2n+1}{3}}$.
\end{enumerate}
\end{theorem}

Let $T$ be an orderly spanning tree of $G$, where $v_1,v_2,\ldots,v_n$
is the counterclockwise preordering of $T$.  Our floor-planning
algorithm is described as follows. Although the first two steps of our
algorithm follow how Chiang et al.~\cite{ChiangLL01} obtained their
2-visibility drawing of $G$ with respect to $T$, we list them this way
to make the presentation of our algorithm more self-contained.

\begin{figure}
\centerline{\input{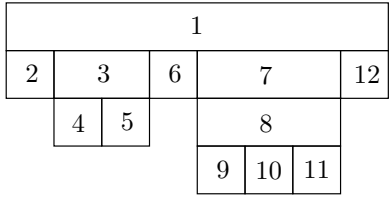}}
\caption{Step 1: visibility drawing of $T$.}
\label{figure:step1}
\end{figure}

\begin{figure}
\centerline{\input{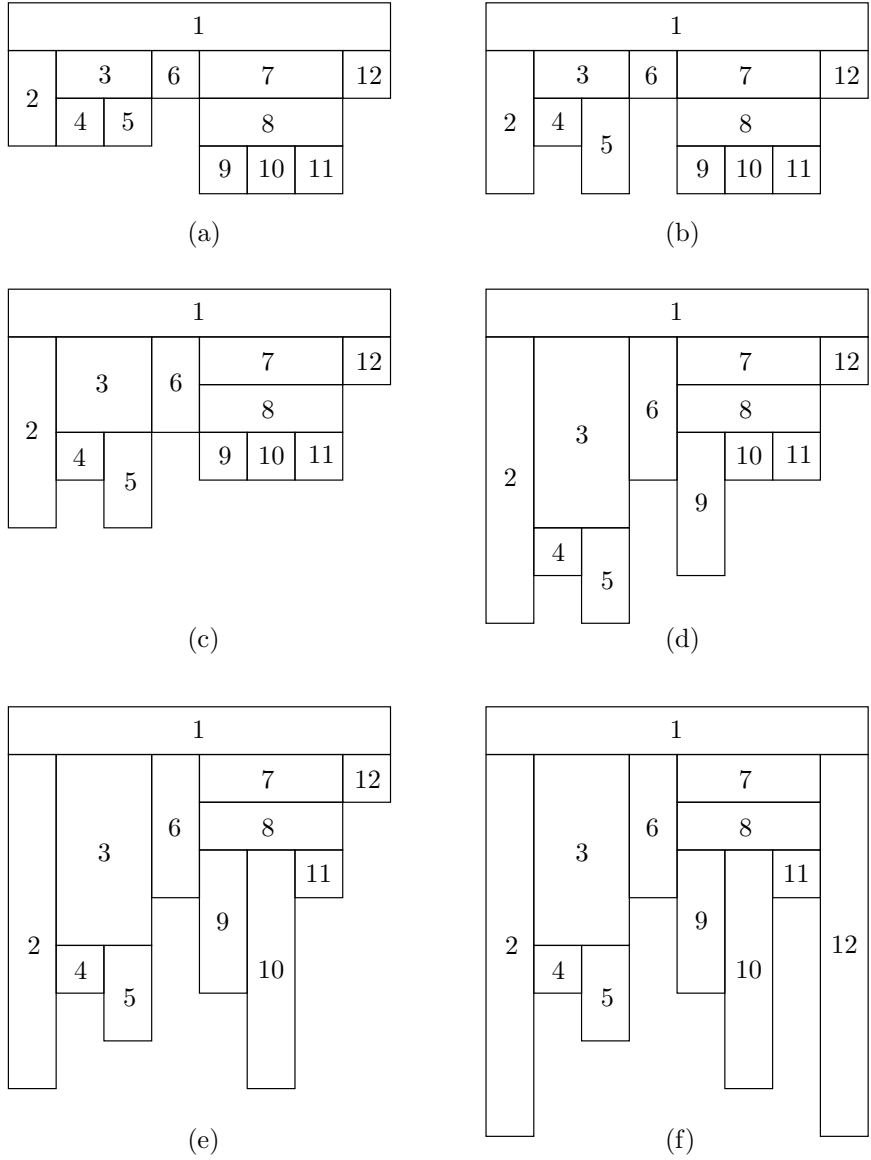}}
\caption{Step 2: obtaining a 2-visibility drawing of $G$ from the
visibility drawing of $T$ by ensuring the horizontal visibility
between $v_i$ and each node in $B_2(v_i)$ for (a) nodes 3 and 4, (b)
node 5, (c) nodes 6--8, (d) node 9, (e) node 10, and (f) nodes 11 and
12.}
\label{figure:step2}
\end{figure}

\begin{figure}
\centerline{\input{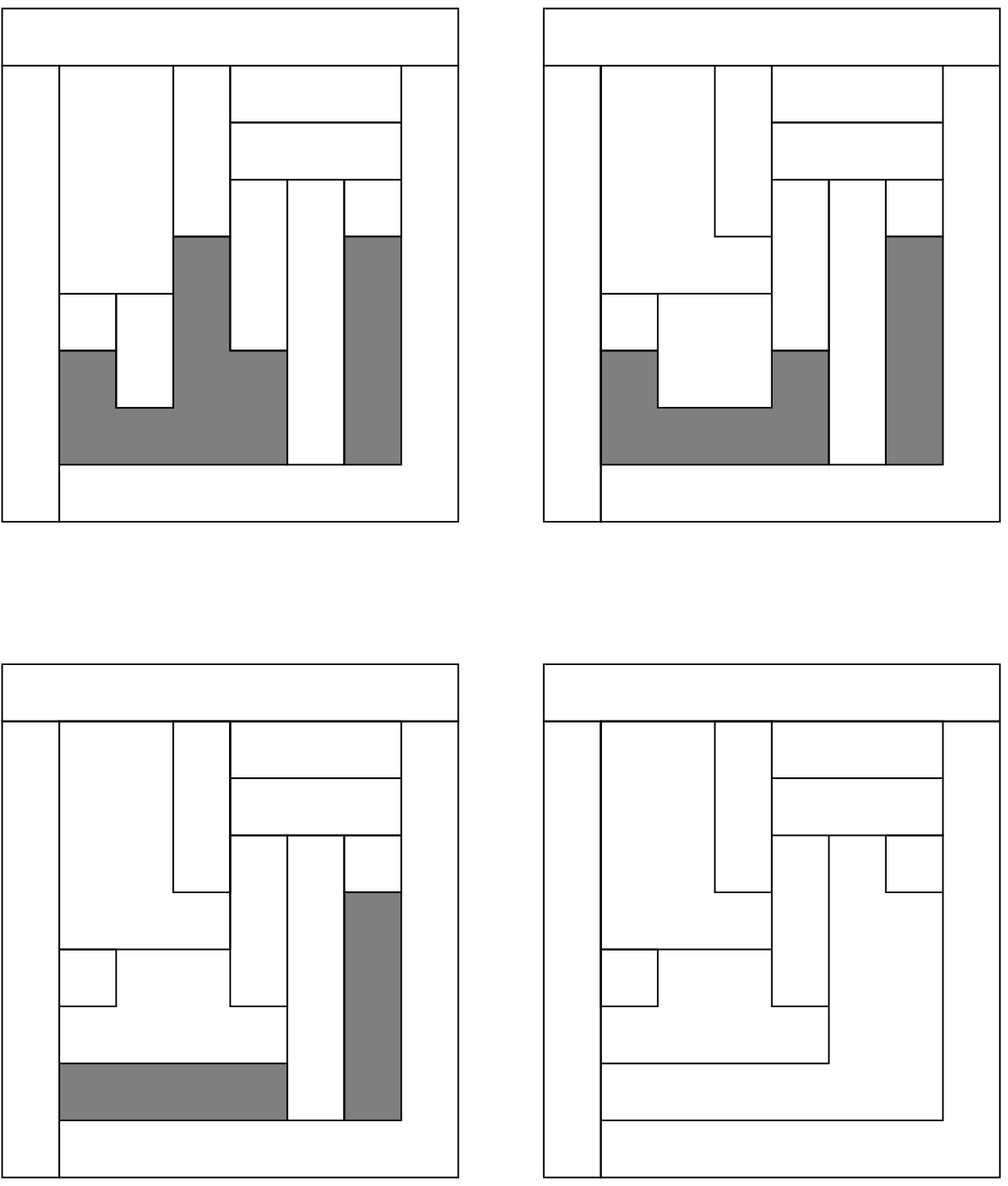}}
\caption{Step 3: growing the horizontal branches for (a) node 12, (b) node 3, (c) nodes 4 and 5, and (d) nodes 6--11.}
\label{figure:step3}
\end{figure}

\medskip
\noindent{\bf Algorithm} $\mbox{\sc FloorPlan}(G,T)$
\begin{description}
\item [Step 1.]  Produce a (vertical) visibility drawing of $T$ as
follows: For each $i=1,2,\ldots, n$, if $v_i$ is a leaf of $T$, then
draw $v_i$ as a unit square; otherwise, draw $v_i$ as a $1\times w(i)$
rectangle. Place each node beneath its parent such that the children
of each node is placed in the same order as in $T$.

\item [Step 2.]  Turn the above visibility drawing of $T$ into a
2-visibility drawing of $G$ by stretching the nodes downward in the
least necessary amount such that $v_i$ and $v_j$ are horizontally
visible to each other if and only if $(v_i,v_j)$ is an unrelated edge
of $G$ with respect to $T$.  Specifically, for each $i=3,4,\ldots,n$,
the $i$-th iteration of this step ensures the horizontal visibility
between $v_i$ and each node in $B_2(v_i)$.

%

\item[Step 3.] First, grow a horizontal branch for $v_n$ from boundary
of $v_n$ visible to $v_2$ such that the left boundary of the
horizontal branch touches $v_2$.  Second, for each
$i=3,4,\ldots,n-1$, grow horizontal branches for $v_i$ from the
boundaries of $v_i$ visible to $v_{\ell(i)}$ and $v_{r(i)}$ such
that the left (respectively, right) boundary of the horizontal
branch touches $v_{\ell(i)}$ (respectively, $v_{r(i)}$).
Furthermore, when extending the boundary of $v_i$, we also extend
the boundaries of the descendants of $v_i$ to maintain the
property that the bottom boundary of each internal node of $T$ is
completely occupied by the top boundaries of its children. Note
that some former extended modules might be covered by latter
extending.

\item[Step 4.] For each $i=n-1,n-2,\ldots,3$, if $v_i$ has a horizontal
branch with height greater than one, then reduce the height of the
thick branch down to one.
\end{description}

Pictures of intermediate steps are
shown to illustrate how our algorithm obtains the floor-plan in
Figure~\ref{figure:example}(b) for the plane graph $G$ with respect to
the orderly spanning tree $T$ shown in Figure~\ref{figure:example}(a).
Figure~\ref{figure:step1} shows how Step~1 obtains the visibility drawing for
$T$.
Figure~\ref{figure:step2} shows how Step~2 
obtains the resulting 2-visibility drawing for $G$.
Observe that the resulting drawing
satisfies the property that the bottom boundary of each internal
node of $T$ is completely occupied by the top boundaries of its
children.
Figure~\ref{figure:step3} illustrates
how Step 3 obtains the resulting drawing for $G$.
Note that when the horizontal
branch of node 3 is extended to the right by one unit to touch the
left boundary of node 9, the right boundary of node 5 is also extended
to the right by the same amount.
%
To see the necessity of Step~4, 
one can verify that the module for node 10 in
Figure~\ref{figure:step3}(d) has a thick horizontal branch. The height
of this thick branch can be reduced by moving down the top boundary of
the thick branch that is adjacent to the bottom boundary of node 11.
The resulting floor-plan consists of only I-modules, L-modules, and
T-modules. Moreover, each horizontal branch of the L-modules and
T-modules has height exactly one.

\begin{lemma}
\label{lemma:main}
The following statements hold for our algorithm {\sc FloorPlan}.
\begin{enumerate}
\item The algorithm can be implemented to run in
$O(n)$ time.
\item The output is a floor-plan of $G$ of size no more than $(n-1)\times w(v_1)$.
\item The resulting floor-plan consists of I-modules, L-modules, and
T-modules, where the height of each horizontal branch of L-modules and
T-modules is one.
\end{enumerate}
\end{lemma}
\begin{proof}
Statement 1. 
One can verify that our algorithm is implementable to run in linear time as follows.
\begin{description}
\item[{\rm Step 1.}] Since $w(v_1),w(v_2),\ldots,w(v_n)$ can be computed from $T$ in
$O(n)$ time, the described (vertical) visibility drawing of $T$ can
easily be computed in $O(n)$ time.

\item[{\rm Step 2.}]  Note that we have to ensure that $v_i$ and $v_j$
are horizontally visible to each other if and only if $v_j\in
B_2(v_i)$ at the end of the stretch-down iteration for
$v_i$. Therefore, when the boundaries of $v_i$ and the nodes in
$B_2(v_i)$ are stretched down, the boundaries of some other nodes
might require being stretched down as well. For example, when we
obtain Figure~\ref{figure:step2}(c) from Figure~\ref{figure:step2}(b)
by stretching down the boundary of node 6 to ensure that nodes 6 and 8
are horizontally visible to each other, we also have to increase the
the heights of nodes 2 and 3 by one.  Thus, a naive implementation of
this step may require $\Omega(n^2)$ time.  However, this step can be
implemented by directly computing the position $y(i)$ of the bottom
boundary of $v_i$ for each $i=1,2,\ldots,n$ and the position $y(i,j)$
of the bottom boundary of each unrelated edge $(v_i,v_j)$ with $i<j$
according to the following recurrence relation:
\begin{eqnarray*}
y(i)&=&\left\{
\begin{array}{ll}
1&\mbox{if $i=1$};\\
\max\setof{y(\ell(i),i),y(i,r(i))}&\mbox{otherwise};
\end{array}
\right.\\
y(i,j)&=&1+\max\setof{y_\ell(i,j),y_r(i,j)},
\end{eqnarray*}
where $y_\ell(i,j)$ and $y_r(i,j)$ are defined as follows.  Let
$v_{j'}$ be the neighbor of $v_i$ that immediately succeeds $v_j$ in
counterclockwise order around $v_i$.  Let $v_{i'}$ be the neighbor of
$v_j$ that immediately precedes $v_i$ in counterclockwise order around
$v_j$.  By $i<j$, one can easily see that either $i'=p(j)$ or
$v_{i'}\in B_2(v_j)$ holds. Similarly, either $j'=p(i)$ or $v_{j'}\in
B_4(v_i)$ holds.  Let
\begin{eqnarray*}
y_\ell(i,j)&=&\left\{
\begin{array}{ll}
y(j')&\mbox{if $j'=p(i)$};\\
y(i,j')&\mbox{otherwise};
\end{array}
\right.\\
y_r(i,j)&=&\left\{
\begin{array}{ll}
y(i')&\mbox{if $i'=p(j)$};\\
y(i',j)&\mbox{otherwise}.
\end{array}
\right.
\end{eqnarray*}
Clearly, the bottom positions $y(i)$ of all nodes $v_i$ can be
obtained in $O(n)$ time by dynamic programming. Since the top position
of $v_i$ is simply $y(p(i))$, the resulting 2-visibility drawing of
$G$ can be obtained in $O(n)$ time.

\item[{\rm Step 3.}]  On the one hand, a naive implementation of this
step may require $\Omega(n^2)$ time, since growing the horizontal
branches for a node may cause boundary extension for its
descendants. On the other hand, although in the $i$-th iteration we
are supposed to extend the boundary of some descendants $v_j$ of
$v_i$, we do not need to actually extend the boundaries of $v_j$ until
the beginning of the $j$-th iteration. Note that how far should the
boundary of $v_j$ be extended can be determined directly from the
boundary of $v_{p(j)}$ in the $j$-th iteration. Clearly, the above
``lazy'' strategy reduces the time complexity of this step to $O(n)$.
Since the unrelated edge $(v_i,v_j)$ of $G-T$ with $i<j$ and
$(v_i,v_j)\ne(v_2,v_n)$ satisfies exactly one equality of $i=\ell(j)$
and $j=r(i)$, the
resulting drawing is a partition of a rectangle into $n$ rectilinear
regions. (That is, there is no gap among modules in the rectangle.)
To prove that the resulting drawing is indeed a floor-plan of $G$, it
suffices to show that growing a horizontal branch of $v_i$ is to reach
the boundary of $v_j$ does not result in new adjacency among these
rectilinear modules. Suppose $v_k$ is a node whose bottom boundary
touches the top bottom of the horizontal branch of $v_i$. Assume for a
contradiction that $v_k$ is not adjacent to $v_i$ in $G$. Since the
resulting drawing of the previous step is a 2-visibility drawing of
$G$, there must be a node $v_{k'}$ lies between $v_i$ and $v_k$
preventing their horizontal visibility to each other. It follows that
there is a face of $G$ containing at least four nodes
$v_i,v_j,v_k,v_{k'}$, contradicting the fact that $G$ is triangulated.

\item[{\rm Step 4.}]
Since $T$ is an orderly spanning tree of $G$ and $G$ is a plane
triangulation, one can see that if $v_i$ grows a horizontal
branch to reach $v_j$, then there must be a unique node $v_k$
whose bottom boundary touches the top boundary of that horizontal
branch of $v_i$. It is also not difficult to verify that both
$(v_i,v_k)$ and $(v_j,v_k)$ are unrelated edges $G$ with respect
to $T$. Thus, in the resulting drawing of the previous step, the
left and right boundaries of $v_k$ have to touch $v_i$ and $v_j$.
Therefore, the height of that horizontal branch of $v_i$ can be
reduced to one by moving downward the bottom boundary of $v_k$,
which is also the top boundary of that horizontal branch, without
changing the adjacency of $v_k$ to other nodes in the floor-plan.
Clearly, each height-reducing operation takes $O(1)$ time by
adapting lazy strategy, so this step runs in $O(n)$ time.  Since
the for-loop of this step proceeds from $i=n-1$ down to $3$, each
horizontal branch has height exactly one at the end of this step.
\end{description}

Statement 2. 
Since Steps~3 and~4 do not affect the adjacency among the rectilinear
modules, it suffices to ensure that the 2-visibility drawing obtained
in Step 2 has size no more than $(n-1)\times w(v_1)$. By the
definition of Steps 1 and~2, it is straightforward to see that the
width of the resulting drawing is precisely $w(v_1)$. It remains to
show that $y(2,n)$, which is exactly the height of the resulting
2-visibility drawing, is no more than $n-1$ as follows. Assume for a
contradiction that $y(2,n)\geq n$. It follows that there is a sequence
of unrelated edges $(v_{s_1},v_{t_1}), (v_{s_2},v_{t_2}),\ldots,
(v_{s_n},v_{t_n})$ with
\begin{displaymath}
2=s_n\leq s_{n-1}\leq s_{n-2}\leq \cdots \leq s_1 < t_1 \leq t_2 \leq \cdots
\leq t_n=n
\end{displaymath}
such that at least one of $s_{i}\ne s_{i+1}$ and $t_{i}\ne t_{i+1}$
holds for each $i=1,2,\ldots,n-1$. It follows that the set
$\setof{s_1,s_2,\ldots,s_n,t_1,t_2,\ldots,t_n}$ contains at least $n$
distinct integers, thereby, contradicting the assumption $2\leq
s_i,t_i\leq n$.

Statement 3. By the definition of Step 3, one can easily verify that
the resulting floor-plan consists of I-modules, L-modules, and
T-modules. By the height-reducing operation performed on the
horizontal branches in Step 4, the statement is proved.
\end{proof}

We are ready to prove the main theorem as follows.

\begin{proof}[Proof for Theorem~\ref{theorem:main}]\quad%
Straightforward by Lemmas~\ref{lemma:orderly} and~\ref{lemma:main}.
\end{proof}

\section{Lower bounds on the worst-case area of floor-plan}
\label{section:4}
This section shows the near optimality of the output of our algorithm.
\begin{lemma}
For each $n\geq 3$, there
is an $n$-node plane triangulation graph $G_n$ 
such that any $h_n\times w_n$ floor-plan of $G_n$ satisfies
$\min\setof{h_n,w_n}\geq \floor{\frac{2n+1}{3}}$ and $h_n+w_n\geq
\ceiling{\frac{4n}{3}}$.
\end{lemma}
\begin{proof}
The lower-bound examples are constructed inductively: For each $n\geq
4$, $G_n$ is obtained from $G_{n-3}$ by adding an external triangle
and arbitrarily triangulating the face between the external triangle
of $G_n$ and the external boundary of $G_{n-3}$. As for the base
cases, let $G_n$ be an arbitrary $n$-node plane triangulation for each
$n=3,4,5$.  Now we show that the required inequalities hold for
each $n\geq 3$.  As for the inductive basis, one can verify
$\min\setof{h_3,w_3}\geq 2$, $h_3+w_3\geq 4$, $\min\setof{h_4,w_4}\geq
3$, $h_4+w_4\geq 6$, $\min\setof{h_5,w_5}\geq 3$, and $h_5+w_5\geq 7$.
Therefore the inequalities hold for the base cases. It remains to ensure
the induction step as follows.
\begin{eqnarray*}
\min\setof{h_n,w_n}&\geq&\min\setof{h_{n-3},w_{n-3}}+2\\
     &\geq&\floor{\frac{2(n-3)+1}{3}}+2\\
     &=&\floor{\frac{2n+1}{3}};
\end{eqnarray*}
\begin{eqnarray*}
h_n+w_n&\geq&h_{n-3}+w_{n-3}+4\\
       &\geq&\ceiling{\frac{4(n-3)}{3}}+4\\
       &=&\ceiling{\frac{4n}{3}}.
\end{eqnarray*}
\end{proof}

\section{Conclusion}
\label{section:5}

A linear-time algorithm for producing compact floor-plans for
plane triangulations has been designed. Our algorithm is based
upon a newly developed technique of orderly spanning trees with
bounded number of leaves~\cite{ChiangLL01}.  In comparison with
previous work on floor-planning plane triangulations~\cite{He99},
our algorithm is simpler in the algorithm itself as well as in
the resulting floor-plan in the sense that the Z-modules required
by~\cite{He99} is not needed in our design.  Another important
feature of our algorithm is the upper bound $(n-1)\times
\floor{\frac{2n+1}{3}}$ on the area of the output floor-plan.
Previous work~\cite{He99,YeapS93} does not provide any area
bounds on their outputs.  Investigating whether the $(n-1)\times
\floor{\frac{2n+1}{3}}$ area is worst-case optimal is an
interesting future research direction.

\section*{Acknowledgment}
We thank the anonymous referees for their helpful comments, which
significantly improve the presentation of the paper.  We also thank
Ho-Lin Chen for his comments regarding an early version of this work.

\bibliographystyle{elsart-num}
\bibliography{floor}

\begin{thebibliography}{10}
\expandafter\ifx\csname url\endcsname\relax
  \def\url#1{\texttt{#1}}\fi
\expandafter\ifx\csname urlprefix\endcsname\relax\def\urlprefix{URL }\fi

\bibitem{TsukiyamaKS86}
S.~Tsukiyama, K.~Koike, I.~Shirakawa, An algorithm to eliminate all complex
  triangles in a maximal planar graph for use in {VLSI} floorplan, in:
  Proceedings of the IEEE International Symposium on Circuits and Systems,
  1986, pp. 321--324.

\bibitem{MailingMH82}
K.~Mailing, S.~H. Mueller, W.~R. Heller, On finding most optimal rectangular
  package plans, in: Proceedings of the 19th Annual IEEE Design Automation
  Conference, 1982, pp. 263--270.

\bibitem{He99}
X.~He, On floor-plan of plane graphs, SIAM Journal on Computing 28~(6) (1999)
  2150--2167.

\bibitem{KozminskiK85}
K.~Ko{\'z}mi{\'n}ski, E.~Kinnen, Rectangular duals of planar graphs, Networks
  15~(2) (1985) 145--157.

\bibitem{KozminskiK98}
K.~A. K{\'o}zmi{\'n}ski, E.~Kinnen, Rectangular dualization and rectangular
  dissections, IEEE Transactions on Circuits and Systems 35~(11) (1988)
  1401--1416.

\bibitem{LaiL90}
Y.~T. Lai, S.~M. Leinwand, A theory of rectangular dual graphs, Algorithmica
  5~(4) (1990) 467--483.

\bibitem{YeapS93}
K.-H. Yeap, M.~Sarrafzadeh, Floor-planning by graph dualization: $2$-concave
  rectilinear modules, SIAM Journal on Computing 22~(3) (1993) 500--526.

\bibitem{DeFPP90}
H.~{de~Fraysseix}, J.~Pach, R.~Pollack, How to draw a planar graph on a grid,
  Combinatorica 10 (1990) 41--51.

\bibitem{Kant96}
G.~Kant, Drawing planar graphs using the canonical ordering, Algorithmica
  16~(1) (1996) 4--32.

\bibitem{KantH97}
G.~Kant, X.~He, Regular edge labeling of $4$-connected plane graphs and its
  applications in graph drawing problems, Theoretical Computer Science
  172~(1-2) (1997) 175--193.

\bibitem{BhaskerS87}
J.~Bhasker, S.~Sahni, A linear algorithm to check for the existence of a
  rectangular dual of a planar triangulated graph, Networks 17 (1987) 307--317.

\bibitem{BhaskerS88}
J.~Bhasker, S.~Sahni, A linear algorithm to find a rectangular dual of a planar
  triangulated graph, Algorithmica 3 (1988) 247--278.

\bibitem{He93}
X.~He, On finding the rectangular duals of planar triangular graphs, SIAM
  Journal on Computing 22 (1993) 1218--1226.

\bibitem{ChiangLL01}
Y.-T. Chiang, C.-C. Lin, H.-I. Lu, Orderly spanning trees with applications to
  graph encoding and graph drawing, in: Proceedings of the 12th Annual ACM-SIAM
  Symposium on Discrete Algorithms, Washington, D. C., USA, 2001, pp. 506--515.

\bibitem{Schnyder90}
W.~Schnyder, Embedding planar graphs on the grid, in: Proceedings of the First
  Annual {ACM}-{SIAM} Symposium on Discrete Algorithms, 1990, pp. 138--148.

\bibitem{Schnyder89}
W.~Schnyder, Planar graphs and poset dimension, Order 5 (1989) 323--343.

\bibitem{Jacobson89}
G.~Jacobson, Space-efficient static trees and graphs, in: Proceedings of the
  30th Annual Symposium on Foundations of Computer Science, IEEE, Research
  Triangle Park, North Carolina, 1989, pp. 549--554.

\bibitem{MR97}
J.~I. Munro, V.~Raman, Succinct representation of balanced parentheses, static
  trees and planar graphs, in: Proceedings of the 38th Annual Symposium on
  Foundations of Computer Science, IEEE, Miami Beach, Florida, 1997, pp.
  118--126.

\bibitem{ChuangGHKL98}
R.~C.-N. Chuang, A.~Garg, X.~He, M.-Y. Kao, H.-I. Lu, Compact encodings of
  planar graphs via canonical ordering and multiple parentheses, in: K.~G.
  Larsen, S.~Skyum, G.~Winskel (Eds.), Proceedings of the 25th International
  Colloquium on Automata, Languages, and Programming, Lecture Notes in Computer
  Science 1443, Springer-Verlag, Aalborg, Denmark, 1998, pp. 118--129.

\bibitem{Foessmeierkk97}
U.~F\"{o}\ss{}meier, G.~Kant, M.~Kaufmann, 2-visibility drawings of planar
  graphs, in: S.~North (Ed.), Proceedings of the 4th International Symposium on
  Graph Drawing, Lecture Notes in Computer Science 1190, Springer-Verlag,
  California, USA, 1996, pp. 155--168.

\bibitem{Bonichon00}
N.~Bonichon, B.~{Le~Sa\"{e}c}, M.~Mosbah, Orthogonal drawings based on the
  stratification of planar graphs, Tech. Rep. RR--1246--00, Laboratoire
  Bordelais de Recherche en Informatique (LaBRI), presented at the 6th
  International Conference on Graph Theory, Marseille, France, August 28 --
  September 1, 2000 (2000).

\bibitem{SunS93}
Y.~Sun, M.~Sarrafzadeh, Floor-planning by graph dualization: ${L}$-shaped
  modules, Algorithmica 10~(6) (1993) 429--456.

\bibitem{DeFraysseixOR94}
H.~{de~Fraysseix}, P.~{Ossona~de~Mendez}, P.~Rosenstiehl, On triangle contact
  graphs, Combinatorics, Probability and Computing 3 (1994) 233--246.

\end{thebibliography}
\end{document}